\documentclass[12pt]{article}

\input epsf
\usepackage{amssymb,wrapfig,amsmath}
\usepackage{xypic,multirow}
\usepackage[matrix,arrow,curve]{xy}
\usepackage{epsfig}

\begin{document}
\addtolength{\baselineskip}{.5mm}
\newlength{\extraspace}
\setlength{\extraspace}{1.5mm}
\newlength{\extraspaces}
\setlength{\extraspaces}{2mm}

\def\nn{{\cal N}}
\def\rr {{\Bbb R}}
\def\cc {{\Bbb C}}
\def\C {{\mathbb C}}
\def\pp {{\Bbb P}}
\def\zz {{\Bbb Z}}
\def\del {\partial}
\def\cy {Calabi--Yau}
\def\ka {K\"ahler}
\def\be{\begin{equation}}
\def\ee{\end{equation}}
\def\vp{\varphi}
\def\oo{\mathcal{O}}
\newcommand{\inv}[1]{{#1}^{-1}} 
\newcommand{\Section}[1]{\section{#1} \setcounter{equation}{0}}
\makeatletter
\@addtoreset{equation}{section}
\makeatother
\renewcommand{\theequation}{\thesection.\arabic{equation}}

\begin{center}
{\hfill RUNHETC-2008-11\\}

\medskip

{\Large \bf{Wall Crossing of BPS States on the Conifold from Seiberg Duality and Pyramid Partitions}}

\bigskip

Wu-yen Chuang and Daniel Louis Jafferis
\medskip

{\it NHETC, Department of Physics, Rutgers University}

{\it 126 Frelinghuysen Rd, New Jersey 08854, USA}

{\it wychuang@physics.rutgers.edu, jafferis@physics.rutgers.edu}

\bigskip

{\bf Abstract}
\end{center}

In this paper we study the relation between pyramid
partitions with a general empty room configuration (ERC) and the
BPS states of D-branes on the resolved conifold. We find that the
generating function for pyramid partitions with a length $n$ ERC
is exactly the same as the D6/D2/D0 BPS partition function on the
resolved conifold in particular K\"ahler chambers. We define a new
type of pyramid partition with a finite ERC that counts the BPS
degeneracies in certain other chambers.

The D6/D2/D0 partition functions in different chambers were
obtained by applying the wall crossing formula. On the other hand,
the pyramid partitions describe $T^3$ fixed points of the moduli
space of a quiver quantum mechanics. This quiver arises after we
apply Seiberg dualities to the D6/D2/D0 system on the conifold and
choose a particular set of FI parameters. The arrow structure of
the dual quiver is confirmed by computation of the Ext group
between the sheaves. We show that the superpotential and the
stability condition of the dual quiver with this choice of the FI
parameters give rise to the rules specifying pyramid partitions
with length $n$ ERC.

\section{Introduction}

One of the most fruitful areas of overlap between string theory
and mathematics has been in the applications of topological field
theories and string theories to questions involving integration
over various moduli spaces of interesting geometrical objects. One
example is that the topological sector of the worldvolume theory
of a D6 brane wrapping a Calabi-Yau 3-fold has been identified
with Donaldson-Thomas theory. The bound states of D2 and D0 branes
to a D6 brane can be regarded as instantons in the topologically
twisted ${\cal N} = 2$ $U(1)$ Yang-Mills theory in six dimensions.
These instantons turn out to correspond to ideal sheaves on the
3-fold. Moreover, Donaldson-Thomas theory involves integration of
a virtual fundamental class over the moduli space of such ideal
sheaves.

There has recently been tremendous progress in understanding the
K\"ahler moduli dependence of the index of BPS bound states of
D-branes wrapping cycles in a Calabi-Yau 3-fold \cite{DM}. On the
mathematical side, various generalizations of Donaldson-Thomas
theory have been proposed \cite{KS}, in which ideal
sheaves are replaced by more general stable objects in the derived
category. These invariants will thus also have a dependence on the
background K\"ahler moduli, as encoded in a stability condition.

Much work has been done on Donaldson-Thomas theory in toric
Calabi-Yau manifolds, which are of necessity noncompact. In
particular, using toric localizations, the theory can be solved
exactly \cite{qfoam, MNOP}. In this work we will focus on the
particular example of the resolved conifold. More recently, there
was some extremely interesting work \cite{Szendroi, BenYoung}, in
which the Donaldson-Thomas invariants on a noncommutative
resolution of the conifold were determined. In that case, the
torus fixed points of the moduli space of noncommutative ideal
sheaves were identified with pyramid partitions in a length $1$
ERC.

In \cite{JM}, these results were reproduced using the physical
techniques resulting from the supergravity description of such
bound states as multi-centered black holes in IIA string theory
compactified on a Calabi-Yau manifold. The Donaldson-Thomas
partition function of the commutative and noncommutative
resolutions of the conifold were shown to arise as special cases
in the moduli space of asymptotic K\"ahler parameters. Moreover,
the D6/D2/D0 partition function was determined in all chambers of
a certain real three parameter moduli space that captured the
relevant universal behavior as a compact Calabi-Yau threefold
degenerated to the noncompact resolved conifold\footnote{These
partition functions were also derived first in \cite{Naka} using a
different approach.}.

In this work, we will demonstrate an intriguing relationship
between the pyramid partitions in a length $n$ ERC and the
D6/D2/D0 partition function in various chambers. In other
chambers, we will give evidence that the torus fixed points of the
moduli space of BPS states are in one to one correspondence with a
new type of pyramid partitions in a finite region.

It is possible to find a basis of D-brane charges which is both
primitive (ie. it generates the entire lattice) and rigid (ie. the
basis branes have no moduli) for D6/D2/D0 branes in the resolved
conifold. Note that this is never possible in compact Calabi-Yau.
Thus the BPS bound states are completely described by the
topological quiver quantum mechanics, whose fields are the open
strings stretched between the basis branes. The quiver that
describes D2/D0 in the conifold is in fact the famous $U(N) \times
U(M)$ Klebanov-Witten quiver \cite{KW}, viewed as a 0-dimensional
theory in our context.

The new ingredient that we introduce is an extended quiver that
also includes a $U(1)$ node associated to the D6 brane, in analogy
with the case of D6/D0 bound states in $\mathbb{C}^3$ studied in
\cite{Jafferis:2007sg}. We determine the spectrum of bifundamental
strings by computing the appropriate $Ext$ groups. One beautiful
feature of this system is that the $SU(2)\times SU(2)$ symmetry of
the conifold completely fixes the superpotential, $SW$, up to
field redefinition, so we do not need to compute it directly. The
moduli space of vacua of this quiver theory depends on the
background K\"ahler moduli through the Fayet-Iliopoulos
parameters.

This moduli space is obtained by imposing the F-term equations,
$\partial W = 0$, on the K\"ahler quotient of the space of fields
(here a finite dimensional space of matrices with ranks determined
by the D-brane charges) by the $U(N)\times U(M) \times U(1)$ gauge
group. The K\"ahler quotient is equivalent to the quotient by the
complexified gauge group, together with an algebraic stability
condition that depends on the FI parameters \cite{King}.

In each chamber of the K\"ahler moduli space of the resolved
conifold found in \cite{JM}, we identify a pair of primitive
sheaves carrying D2/D0 charge that become mutually BPS at the
boundary of the K\"ahler cone, in that chamber. In particular,
along the boundary of the K\"ahler cone in the chamber $C_n =
[{\cal W}^{-1}_n {\cal W}^{-1}_{n-1}]$, in the notation of
\cite{JM} for $n>0$, the central charges of ${\cal O}_C(-n-1)[1]$
and ${\cal O}_C (-n)$ become aligned. In the chamber $\tilde C_n =
[{\cal W}^1_{n-1} {\cal W}^1_n]$, it is ${\cal O}_C(-n)$ and
${\cal O}_C (-n+1)[1]$ which become mutually BPS at the boundary
of the K\"ahler cone.

Therefore in that chamber, we choose those sheaves, together with
the pure D6 brane, ${\cal O}_X$, as our basis objects, and
construct the quiver theory, as explained above. Note that all of
the above pairs for different $n$ are related to each other by
application of Seiberg duality. We will denote the quivers
resulting from the choice of basis branes above as $Q_n$ and
$\tilde Q_n$ respectively. The fact that the central charges
become aligned in that chamber implies that the bifundamental
strings between them are massless at tree level, hence the FI
parameters, $\theta$, for the two D2/D0 nodes must be equal for
those values of the K\"ahler moduli.

It is easy to check that $\theta_{{\cal O}_X[1]} > 0$,
$\theta_{{\cal O}_C(-n-1)[1]} < 0$, and $\theta_{{\cal O}_C(-n)} <
0$ is a single chamber, and includes the locus where $\theta_{{\cal
O}_C(-n-1)[1]} = \theta_{{\cal O}_C(-n)}$. Therefore we can
identify that chamber in the space of FI parameters for the quiver
$Q_n$ with the chamber $C_n$ in the space of background K\"ahler
moduli.

We show that with that choice of FI parameters, the King stability
condition becomes equivalent to a simple cyclicity - the quiver
representations must be generated by a vector in the
$\mathbb{C}^1$ associated to the D6 brane $U(1)$ node. The
relations obtained from the superpotential are used to see that
the torus fixed points of the moduli space of stable
representations in this chamber are exactly the pyramid partitions
in a length $n$ ERC defined in \cite{Szendroi}! Similarly, for the
chambers $\tilde C_n$, we demonstrate that the torus fixed points
are in one to one correspondence with pyramid partitions in a
certain finite empty room configuration that we introduce.

The generating functions for pyramid partition in a length $n$ ERC
was determined in \cite{Szendroi, BenYoung}, and we check that it
agrees with the D6/D2/D0 partition function in the resolved
conifold in the appropriate chamber found in \cite{JM} \cite{Naka} \cite{Nagao}. This
requires correctly changing variables to take into account the
D2/D0 charges of the basis sheaves used in the construction of the
quivers. We explicitly check a few examples of the finite type
pyramid partitions as well.

Thus we have been able to reproduce the D6/D2/D0 partition
function on the resolved conifold, in a given chamber in the space
of background K\"ahler moduli, by judiciously choosing a
particular Seiberg dual version of the associated three node
quiver in which the FI parameters corresponding to the K\"ahler
moduli are of a special simple form.

The paper is organized as follows. In section $2$ we give a review
of wall crossing formulae and the D6/D2/D0 partition function on
the resolved conifold based on \cite{DM,JM}. In section $3$, we
introduce the relation between the pyramid partition function and
the D6/D2/D0 BPS partition function. In section $4$ we  derive the
Seiberg dual quiver and then compute its superpotentials and arrow
structures. Next the discussion is on the stability condition and
the rules of pyramid partitions. Some future directions will be
presented in the conclusion part.

\bigskip

\section{Wall crossing formula and D6/D2/D0 BPS partition function on the resolved conifold}
The index of BPS states with a given total charge is an integer,
and thus is a piecewise constant function of the background values
of the K\"ahler moduli. Moreover, the fact that it is a
supersymmetric index implies that it can only jump when a state
goes to infinity in the moduli space of BPS states, that is when
the asymptotics of the potential change. The only known way this
can happen for the case of BPS bound states of D-branes wrapping a
Calabi-Yau manifold is that the physical size of a multi-centered
Denef black hole solution diverges at some value of the K\"ahler
parameters \cite{Denef1}. This occurs exactly at (real codimension 1) walls of
marginal stability, when the central charges, $Z_1$ and $Z_2$, of
the two constituents, $\Gamma_1$ and $\Gamma_2$, of the
multi-centered supergravity solution become aligned.

At such a wall of marginal stability $t=t_{ms}$ corresponding to a
decay $\Gamma \to \Gamma_1 + \Gamma_2$, the BPS index will have a
discrete jump given by

\be \Delta \Omega (\Gamma, t) = (-1)^{\langle\Gamma_1,
\Gamma_2\rangle-1} \ |\langle\Gamma_1,\Gamma_2\rangle | \
\Omega(\Gamma_1,t_{ms})\Omega(\Gamma_2, t_{ms}), \ee where
$\Gamma_1$ and $\Gamma_2$ are primitive. A semi-primitive wall
crossing formula is also given in \cite{DM}.

\be \Omega(\Gamma_1) + \sum_{N} \Delta \Omega(\Gamma_1 + N
\Gamma_2) q^{N} = \Omega(\Gamma_1) \prod_{k>0}(1-(-1)^{k\langle
\Gamma_1, \Gamma_2\rangle}q^k)^ {k | \langle\Gamma_1,\Gamma_2
\rangle| \Omega(k \Gamma_2)}. \ee

This formula gives a powerful way to construct the D6/D2/D0 BPS
generating function on the resolved conifold from the
Donaldson-Thomas generating function \cite{JM}. The absence of
higher genus Gopakumar-Vafa invariants in the resolved conifold
implies that only the pure D6 brane exists as a single centered
solution. Thus in the core region of the K\"ahler moduli space,
the D6/D2/D0 partition function is just $Z=1$.

The position of the relevant walls of marginal stability was
determined in \cite{JM}, and using the wall crossing formula for a
single D6 bound arbitrary numbers of D2/D0 fragments, the
partition function was then computed throughout the moduli space.
The wall of marginal stability for $\Gamma = 1 - m' \beta + n' dV$
with $\Gamma_h = -m_h \beta + n_h dV$ in a compact Calabi-Yau
manifold, $X$, was shown have a well defined limit as the geometry
approached that of the noncompact resolved conifold. Moreover, the
walls are independent of $m'$ and $n'$, and separate chambers in a
real three dimensional space parameterized by the K\"ahler size,
$z$, of the local $\mathbb{P}^1$, and a real variable $\vp =
\frac{1}{3} \arg(Vol_X)$ that characterizes the strength of the
B-field along the noncompact directions in units of the K\"ahler
form.

The wall of marginal stability for the fragment $\Gamma_h$ was
denoted by ${\cal W}^{m_h}_{n_h}$. The final result for the index
of D6/D2/D0 bound states found in \cite{JM} was that in the
chamber between ${\cal W}^1_n$ and ${\cal W}^1_{n+1}$, the
generating function is \begin{equation}\label{part+} Z(u,v; [{\cal
W}^1_n {\cal W}^1_{n+1}]) = \prod_{j=1}^n \left(1 - (-u)^j
v\right)^j.
\end{equation} Similarly, in the chambers where negative D2 charges
appear, \begin{equation}\label{part-} Z(u,v; [{\cal W}^{-1}_{n+1}
{\cal W}^{-1}_n]) = \prod_{j > 0} \left( 1 - (-u)^j \right)^{-2j}
\left(1 - (-u)^j v \right)^j \prod_{k > n} \left(1 - (-u)^k v^{-1}
\right)^k.
\end{equation} In the extreme case, $n=0$, of the latter they found agreement
with the results of Szendr\H{o}i, who calculated the same
partition function at the conifold point using equivariant
techniques to find the Euler character of a moduli space of
noncommutative sheaves. The $n \rightarrow \infty$ limit of
(\ref{part-}), one obtains the usual ``large radius''
Donaldson-Thomas theory that was determined in \cite{qfoam}, again
using equivariant localization.

\section{Pyramid partition and BPS partition function on resolved conifold}
In this section we will first discuss the relation between the
pyramid partition generating function with length n empty room
configuration (ERC) and the D6/D2/D0 BPS state partition function.
Afterwards we will discuss how the pyramid partition arises when
we look at the torus fixed points on the moduli space
$\mathcal{M}_v$ of representations of the Calabi-Yau algebra $A$
for the conifold quiver.

\subsection{Pyramid partition generating function}

\begin{figure}[ht]
\centering 
\epsfig{file=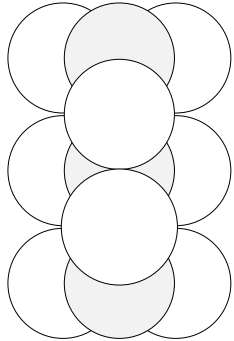, width=3cm}
\caption{This figure illustrates the length 2 ERC of the pyramid partition,
from the zeroth layer to the 2nd layer.}
\label{fig:pyramid}
\end{figure}

The use of pyramid partitions in this context first arose in
\cite{Szendroi}. Consider the arrangement of stones of two
different colors (white and grey) as in Figure \ref{fig:pyramid}.
For a generic ERC with length $n$, there will be $n$ white stones
on the zeroth layer.\footnote{We count the layers as the zeroth,
1st, 2nd, and so on.} On layer $2i$, there are $(n+i)(1+i)$ white
stones, while on layer $2i+1$ there are $(n+i+1)(1+i)$ grey
stones. When we write generating functions, the number of white
stones will be counted the power of $q_0$, and the grey stones by
$q_1$.

A finite subset $\Pi$ of the ERC is a
\textit{pyramid partition}, if for every stone in $\Pi$, the
stones directly above it are also in $\Pi$. Denote as $w_0$ and
$w_1$ the number of white and grey stones in the partition. Also
denote by $\mathcal{P}_n$ the set of all possible pyramid
partitions for the ERC of length $n$.

The generating function is defined combinatorially as

\be Z_{pyramid} (n;q_0,q_1) = \sum_{\Pi \in \mathcal{P}_n}
q_0^{w_0} (-q_1)^{w_1}. \ee

This function can be computed by some dimer shuffling techniques;
we refer the interested reader to \cite{BenYoung} for details.
Here we just quote the result for the generating function for
general $n$ ERC,

\begin{eqnarray}
Z_{pyramid} (n;q_0,q_1) = M(1,-q_0 q_1)^2 \prod_{k \geq 1} (1+q_0^k (-q_1)^{k-1})^{k+n-1} \nonumber \\
\prod_{k \geq 1}(1+q_0^k (-q_1)^{k+1})^{max(k-n+1,0)},
\end{eqnarray} where $M(x,q)$ is the MacMahon function

\be M(x,q) = \prod_{n=1} (\frac{1}{1-xq^n})^n . \ee Notice that
the exponents of the two terms in the product formula start from
$n$ and $1$ respectively. This function turns out to be exactly
the D6/D2/D0 BPS partition function on the resolved conifold found
in \cite{JM} \cite{Naka} \cite{Nagao} in certain chambers
after performing the following ($n$-dependent) parameter
identifications:

\be
\label{eq:identify}
u = -q_0 q_1\ ; \ v = (-q_0)^{n-1} q_1^n .
\ee

Now we have

\begin{eqnarray}
Z_{pyramid} (n;q_0,q_1)&=&Z_{D6/D2/D0} (u,v; C_n) \nonumber \\
&=& M(1,-u) \prod_{k \geq 1} (1- (-u)^k v)^{k} \prod_{k\geq n} (1-
(-u)^k v^{-1})^{k}.
\end{eqnarray}

The upshot is that the pyramid partition for a general empty room
configuration counts the number of D6/D2/D0 BPS bound states at a
certain value of the background modulus! More precisely, this
chamber sits between the conifold point and the large radius
limit; $C_n =  [{\cal W}^{-1}_n {\cal W}^{-1}_{n-1}]$ in the
notation of \cite{JM}.

We will explain that this is no coincidence, after we perform a
Seiberg duality on the original D6/D2/D0 quiver theory. Moreover,
the rules of specifying a pyramid partition encode the stability
condition for these BPS states.

\medskip

In the chamber $\tilde{C}_n = [{\cal W}^{1}_{n-1} {\cal
W}^{1}_{n}] $, the D6/D2/D0 BPS states partition function is
given by

\be Z_{D6/D2/D0} (u,v, \tilde{C}_n)= \prod_{k
=1}^{n-1} (1-(-u)^{-k} v)^{k}. \ee

\begin{figure}[ht]
\centering 
\epsfig{file=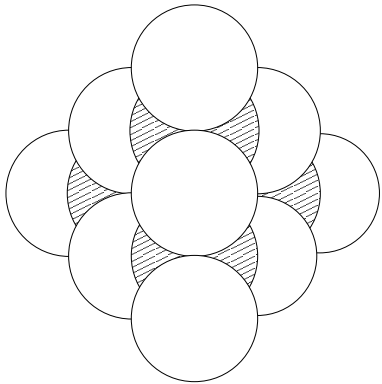, width=5cm}
\caption{This is the ERC for the fintie type pyramid partition with length 3.}
\label{fig:finite}
\end{figure}

We conjecture that the partition
function in these chambers can be described by some \textit{finite
type} pyramid partitions with length $(n-1)$ ERC
(see Figure \ref{fig:finite} for length 3 example), after a change of
variables.

For the finite type pyramid partition with length n, there are $n
\times 1$ white stones on the zeroth layer, $(n-1) \times 1$ grey
stones on the first layer, $(n-1) \times 2$ white stones on the
second, $(n-2) \times 2$ grey stones on the third, and so on until
we reach $1 \times n$. The way of counting is the same. We count
the finite subsets $\Pi$ of the ERC in which,
for every stone in $\Pi$, the stones directly above it are also in
$\Pi$.

The change of variables is given by
\be
u= -q_0 q_1, \ \ \ v=-q_0^nq_1^{n-1}.
\ee

Let us consider length 2 ERC as an example to illustrate the
correspondence between the BPS states partition function in
chambers $\tilde{C}_n$ and the finite type pyramid partition.
\begin{eqnarray}
Z_{n=2 \ ERC}&=&1+2q_0+q_0^2+q_0^2q_1+2q_0^3q_1+q_0^4q_1 \nonumber \\
&=& (1+q_0^2q_1)(1+q_0)^2 = (1-(-u)^{-1}v)(1-(-u)^{-2}v)^2 \nonumber \\
&=& Z(u^{-1},v, \tilde{C}_3).
\end{eqnarray}

\subsection{Conifold quiver and pyramid partitions}

This section is a review of \cite{Szendroi}. Consider the conifold quiver $Q=\{V,E\}$, with two
vertices $V=\{0,1\}$, and four oriented edges $E=\{A_1, A_2: 0 \to 1, B_1, B_2:1 \to 0\}$. The
F-term relations come from the quartic superpotential $W=A_1B_1A_2B_2-A_1B_2A_2B_1$.\cite{KW}

The quiver algebra $A$ contains the idempotent ring
$\mathbb{C}[f_0, f_1]$ and can be given by generators and
relations as \be A=\mathbb{C}[f_0, f_1] \langle A_1,A_2,B_1,B_2
\rangle / \langle B_1A_iB_2-B_2A_iB_1, A_1B_iA_2-A_2B_iA_1, i=1,2
\rangle . \ee $A$ is a smooth Calabi-Yau algebra of dimension
three \cite{Ginzburg} and a crepant non-commutative resolution of
the singularity
$\text{Spec}\left(\mathbb{C}[x_1,x_2,x_3,x_4]/(x_1x_2-x_3x_4)\right)$.

Consider the rank two torus action $T_W$ on the moduli space
$\mathcal{M}_V$ of framed cyclic A-modules \footnote{The torus
action fixing the superpotential is a rank three torus $T_{FW}$,
described by $(A_1, A_2, B_1, B_2) \to (\mu^a A_1, \mu^b A_2,
\mu^c B_1, \mu^{-a-b-c} B_2)$. And $T_W$ is the quotient of
$T_{FW}$ by the $C^*$ action $(\mu, \mu, \mu^{-1},\mu^{-1})$.}. It
has been shown by Szendr\H{o}i \cite{Szendroi} that the
$T_W$-fixed points on the moduli space $\mathcal{M}_V$ are all
isolated and have a one-to-one correspondence with pyramid
partitions $\Pi \in \mathcal{P}_1$ of weight $(w_0, w_1)$. This
weight vector is the same as the rank vector of the corresponding
quiver.

Moreover, given a pyramid partition $\Pi \in \mathcal{P}_1$, we
can obtain the precise framed cyclic module $\mathbb{M}_{\pi}$
defined by it from looking at the pyramid partition. First, we
draw $A_1$ and $A_2$ fields in the perpendicular direction out of
the center of the white stones and draw $B_1$ and $B_2$ fields in
the horizontal direction out of grey stones. The superpotential
F-term relations require that we get the same result if we follow
the arrows of opposite directions of the $A_i$ or $B_i$ fields
down to the three lower layers.

\begin{figure}[ht]
\centering 
\epsfig{file=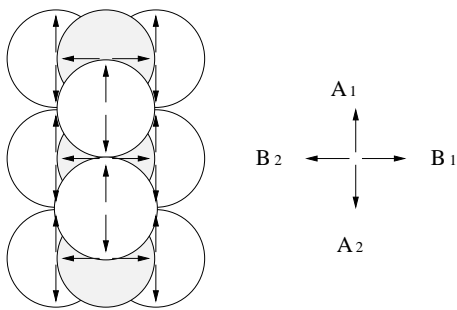, width=6cm}
\caption{This figure illustrates how to define a framed cyclic module from a pyramid partition.}
\label{fig:pyramidarrow}
\end{figure}

The cyclicity property of the module turns into the rule that for
every stone in $\Pi$, the stones directly above it are also in
$\Pi$. The module is generated by the stones on the zeroth layer.
We show in section 5 that the cyclicity condition is equivalent to
the King stability condition in a particular chamber of the quiver
(Fig. \ref{fig:nosheaf}) obtained by introducing a new node for the D6
brane to the Klebanov-Witten quiver discussed in \cite{Szendroi}.

In \cite{Szendroi}, Szendr\H{o}i generalized the notion of pyramid
partitions to length $n$ ERC, and his conjecture for
the resulting generating function was proven in \cite{BenYoung}.
We shall show that these partitions arise as torus fixed points of
the moduli space of King stable representations of the quiver
$Q_n$. For now, note that in the case of general length $n$ ERC,
the $n$ stones on the zeroth layer will still play the roles of
the framing vectors of the quiver and generate the whole module.


There are also new $(n-1)$ relations for $n>1$, if we follow the
arrows from the layer zero to layer one, which read:

\be A_1 q_2 = A_2 q_1, \ A_1 q_3 = A_2 q_2, \ \cdots, A_1 q_n =
A_2 q_{n-1}, \ee where $q_1 \cdots q_n$ are the framing nodes on
the layer zero. Later we will see that these relations arise from
certain cubic terms in the superpotential which are not present
for $n=1$.

\section{Deriving the Quivers via Seiberg Duality}

Recall that a standard choice of the sheaves representing the
conifold quiver is as follows \cite{AK}. \be \mathcal{O}_X[1],\
\mathcal{O}_{C},\ \mathcal{O}_C(-1)[1]. \ee The arrow structure of
this quiver is determined by the Ext group.\footnote{We will use
$\text{Ext}$ and $\underline{Ext}$ to denote the global and local
Ext respectively.}

\be \text{Ext}^1(\mathcal{O}_{C},\mathcal{O}(-1)[1]) \cong
\text{Ext}^1(\mathcal{O}(-1)[1],\mathcal{O}_{C}) \cong
\mathbb{C}^2. \ee

So if we take the rank vector to be $(1,M+N,M)$, the system will
have charges $(D6,D2,D0)=(1,M,N)$. Now we are going to show that
this quiver at certain FI parameters leads to the pyramid
partition after performing Seiberg dualities.

\begin{figure}[ht]
\centering 
\epsfig{file=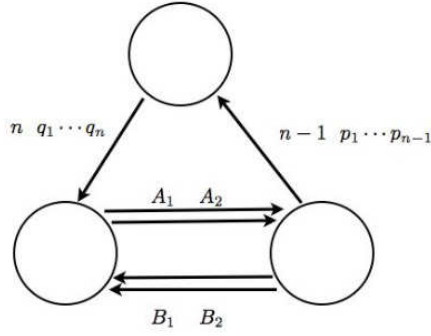, width=6cm}
\caption{The n q fields gives $n$ marked points in the quiver, which generates the whole module, and
the $(n-1)$ p fields will give $(n-1)$ relation via superpotential. The sheaves representing the nodes are
to be determined.}
\label{fig:nosheaf}
\end{figure}

First of all we know that in the pyramid partition there are n marked framing nodes, which are the
most top nodes $q_1 \cdots q_n$  and $n-1$ relations:

\be
\label{eq:qs}
A_1 q_2 = A_2 q_1, \ A_1 q_3 = A_2 q_2, \ \cdots, A_1 q_n = A_2 q_{n-1}.
\ee This implies that we want to find a quiver representation with the
following arrow structure. (See Figure \ref{fig:nosheaf}.)

For this purpose we choose the basis to be \be \mathcal{O}_X,\
\mathcal{O}_C(-n-1),\ \mathcal{O}_C(-n)[-1].\ee

Or equivalently, by an overall shift,\footnote{In order to make contact with the convention in \cite{AK}.}
\be
\mathcal{O}_X[1],\ \mathcal{O}_C(-n-1)[1],\ \mathcal{O}_C(-n): \ \ \ \ \ Q_n.
\ee
The quiver with this basis will be called $Q_n$.

Now let us try to find the sheaves corresponding to the finite type pyramid partition with length $n-1$.
There are $n-1$ framing nodes on the top and $n$ relations coming from the zeroth layer. So what we have to
do is simply to reverse the directions of the p and q fields in Figure \ref{fig:nosheaf}.
The basis of the quiver is given by

\be
\mathcal{O}_X[1],\ \mathcal{O}_C(-n-1),\ \mathcal{O}_C(-n)[1]: \ \ \ \ \ \tilde{Q}_{n-1}.
\ee

\begin{figure}[ht]
\centering 
\epsfig{file=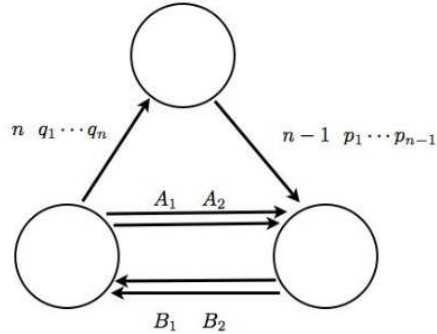, width=6cm}
\caption{This shows the arrow structure of the quiver $\tilde{Q}_{n-1}$.}
\label{fig:nosheaf2}
\end{figure}

In the following section we will summarize the brane charges of
the Seiberg dual quivers $Q_n$ and then confirm the F-term
relations imposed by superpotential and the arrow structures of
the proposed quivers with Ext group computation. The Ext group
computation for the quiver $\tilde{Q}_n$ simply follows from the
computation for $Q_n$; therefore we only focus on $Q_n$ from now
on.
\subsection{Brane charges of the quiver}

The Chern character of the charges of the primitive objects in the
derived category that we used as a basis in the quiver $Q_n$ can
be computed as
\be
\big[ \oo_C(-n)\big] = -\beta + (1-n) d V, \
\ \ \ \ \big[ \oo_C(-n-1)[1] \big] = + \beta + n d V,
\ee in the conventions of \cite{DM}.

In the original D6/D2/D0 system, powers $u$ and $v$ count the D0
and D2 charges respectively. Suppose we have a bound state with
$1$ D6, $M$ D2 and $N$ D0 charges; this will be represented by a
quiver with ranks determined by the following computation:

\be
(-u)^N v^M = (q_0 q_1)^N ((-q_0)^{n-1}q_1^n)^M = (-1)^{n-1} q_0^{N+(n-1)M} q_1^{N+nM}.
\ee Recall that $q_0$ are the number of white stones. Thus the ranks
of the node $\oo_C(-n-1)[1]$ and $\oo_C(-n)$ are $N+(n-1)M$ and
$N+nM$ respectively. This combination indeed gives the right total
charges we are aiming at. We summarize the result in the following
table. (See Table \ref{table:charges}.)

\begin{table}[ht]
\centering 
\begin{tabular}{c c c c} 
\hline\hline 
Sheaves & Ranks & Charge & FI parameters\\ [0.5ex] 
\hline 
$\oo_X[1]$ & 1 & $D6$ or $\bar{D}6$ & $\theta_1$\\
$\oo_C(-n-1)[1]$ & $N+(n-1)M$ & $\bar{D}2,nD0$ & $\theta_2$ \\
$\oo_C(-n)$ & $N+nM$ &  $D2,(n-1)\bar{D}0$ & $\theta_3$\\ [1ex]
\hline
\end{tabular}
\caption{Sheaves, ranks, and charges. Here we have taken into account the induced D0 charge of $\oo_C$.
The total charge of the system is $M$ D2 and $N$ D0.}
\label{table:charges}
\end{table}

\subsection{ $\oo_X \to \oo_C(-n-1)$ }
We now proceed to determine the number of bifundamental fields
that appear in the quiver, by computing the $Ext$ groups between
the basis sheaves. First of all, since $\mathcal{O}_X$ is
projective (thus free), we have

\be
\underline{Ext}^i(\mathcal{O}_X, \mathcal{O}_C)=0, \ \ i>0.
\ee

And we also have $\text{Ext}^0 (\oo_X, \oo_C(-n-1))=0$ when $n \geq 0$ since $\text{Ext}^0 (\oo_X, \ \  )$
is the global section functor. As for $\text{Ext}^1 (\oo_X, \oo_C(-n-1))$, we need to use the following:

\be
\text{dim} \ \text{H}^n ( \mathbb{P}^n, \mathcal{O}(m) ) = \binom{-m-1}{-n-m-1}.
\ee

Therefore,

\be
\text{Ext}^1 (\oo_X, \oo_C(-n-1)) = \text{H}^1 (X, \oo_C(-n-1)) \cong \mathbb{C}^{n},
\ee
\be
\text{Ext}^2 (\oo_X, \oo_C(-n-1))=\text{Ext}^3 (\oo_X, \oo_C(-n-1))=0.
\ee

\subsection{$\oo_C(-n) \lbrack -1 \rbrack \to \oo_X$}

According to Chapter $5.3$ in \cite{GH}, the local sheaves $\underline{Ext}^k(\oo_C, \oo_X)$
are all trivial except for $k=2$.

\be
\underline{Ext}^2(\mathcal{O}_C, \mathcal{O}_X) \cong \iota_{*}\mathcal{K}_C = \iota_{*} \oo_C(-2),
\ee where $\mathcal{K}_C$ is the canonical bundle over $P^1$.

Twisting the sheaf by $\mathcal{O}_X (n)$, we have

\be
\underline{Ext}^2(\mathcal{O}_C (n), \mathcal{O}_X) \cong \iota_{*}\mathcal{K}_C (-n) = \iota_{*} \mathcal{O}_C(-2-n).
\ee

There is a local to global spectral sequence which we can apply to get the Ext group.
However, if $n$ is large, we can simply apply Property $6.9$ in \cite{Hart} to get the Ext group
we want.

The property says, if $\oo_X(1)$ is a very ample invertible sheaf and $\mathcal{E}$ and $\mathcal{F}$ are coherent sheaves on $X$,
there exist an integer $n_0$, depending on $\mathcal{E}$, $\mathcal{F}$ and $i$, such that for $n > n_0$,

\be
\text{Ext}^i(\mathcal{E}, \mathcal{F}(n))= \Gamma(X, \underline{Ext}^i(\mathcal{E}, \mathcal{F}(n))).
\ee

So for $n>>0$, we have
\be
\text{Ext}^1(\oo_C(-n), \oo_X) = \Gamma(X, \underline{Ext}^1(\oo_C, \oo_X(n)))=0,
\ee

\be
\text{Ext}^2(\oo_C(-n), \oo_X) = \Gamma(X, \underline{Ext}^2(\oo_C, \oo_X(n)))= \Gamma(X, \iota_{*}\oo_C(n-2)) \cong \mathbb{C}^{n-1},
\ee

\be
\text{Ext}^1(\oo_C(-n)[-1], \oo_X)  \cong \mathbb{C}^{n-1}.
\ee

\begin{figure}[ht]
\centering 
\epsfig{file=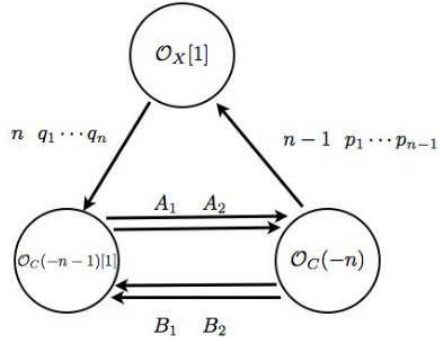, width=6cm}
\caption{The quiver diagram for the pyramid partition with length $n$ ERC.}
\label{fig:sheaf}
\end{figure}

Now we sum up the computation in a quiver diagram, in which we actually apply an overall shift.
(See Figure \ref{fig:sheaf}.)

\subsection{Superpotential}

In principle, the computation of the superpotential of this quiver
quantum mechanics would require evaluating the B-model disk
amplitude with boundary conditions determined given by the basis
B-branes. Luckily that has already been done for the sheaves
$\oo_C(n)$ and $\oo_C(m)[-1]$ by \cite{AK}, resulting in the
Klebanov-Witten superpotential, $$W = Tr (A_1 B_1 A_2 B_2 - A_1
B_2 A_2 B_1).$$ Furthermore, it will turn out that the $SU(2)
\times SU(2)$ symmetry of the resolved conifold will completely
fix the superpotential terms involving the $p$ and $q$ fields, up
to field redefinition.

Recall that the projective coordinates of the resolved conifold,
$$(x_1, x_2, y_1, y_2) \equiv (\lambda x_1, \lambda x_2, \lambda^{-1} y_1, \lambda^{-1} y_2),$$
transform with $(x_1, x_2)$ in the doublet of $SU(2)_1$ and $(y_1,
y_2)$ in the doublet of $SU(2)_2$. In the above derivation of the
$Ext$ groups, it was clear that the $n$ $q$ and $(n-1)$ $p$ fields
live in various cohomology groups of the $\pp^1$. These groups
carry an induced action of the $SU(2)_1$ symmetry, which must be
realized as a global symmetry group of the quiver quantum
mechanics.

Therefore we conclude that $q$ and $p$ are in the $(\bar{n})$ and
$(n-1)$ representation of the global $SU(2)_1$, under which $A_1$
and $A_2$ form a fundamental representation. There is a unique
cubic superpotential that is invariant under this $SU(2)$, up to
field definition, essentially because there is a single copy of
the trivial representation in the tensor product $(\bar{n})
\otimes (n-1) \otimes (2)$. We first can construct from $A$ and
$p$ a combination which is $(n)$ under $SU(2)$. This is basically
the same as constructing angular momentum states $|l+\frac{1}{2},
m+\frac{1}{2} \rangle$ from $|l,m \rangle$ and $|\frac{1}{2}, \pm
\frac{1}{2} \rangle$, where $2l+1=n-1$.

By using the following relation,

\be |l+\frac{1}{2},m+\frac{1}{2}\rangle =
\sqrt{\frac{l+m+1}{2l+1}} |l,m\rangle |\frac{1}{2}, \frac{1}{2}
\rangle +\sqrt{\frac{l-m}{2l+1}} |l,m+1\rangle |\frac{1}{2},
\frac{-1}{2} \rangle , \ee we can write down explicitly the form of superpotential (ignoring
the color trace and index structure and just focusing on the
invariance of the $SU(2)$)

\be W \sim p_1 A_2 q_1 + ( \sqrt{\frac{n-2}{n-1}} p_2 A_2 +
\sqrt{\frac{1}{n-1}}p_1 A_1 ) q_2 + (\sqrt{\frac{n-3}{n-1}} p_3
A_2 + \sqrt{\frac{2}{n-1}}p_2 A_1 ) q_3 + \cdots \ee

We should perform the following field redefinitions:

\be \tilde{q}_1 = q_1 , \ \tilde{q}_2 = - \sqrt{\frac{1}{n-1}}q_2,
\ \tilde{q}_3 = \sqrt{\frac{2}{(n-1)(n-2)}} q_3, \ \tilde{q}_4 = -
\sqrt{\frac{3!}{(n-1)(n-2)(n-3)}} q_4, \ \cdots \ee

The relations implied by superpotential become

\be A_1 \tilde{q}_2 = A_2 \tilde{q}_1, \ A_1 \tilde{q}_3 = A_2
\tilde{q}_2, \ \cdots, A_1 \tilde{q}_n = A_2 \tilde{q}_{n-1}. \ee

Although these field redefinitions are not unitary and will spoil
the D-terms, King's stability condition will not change under
these redefinitions. The moduli space of solutions to the
F-flatness and D-flatness conditions, modded out by the $U(N)
\times U(M)$ gauge symmetry is equivalent to the $GL(N,
\mathbb{C}) \times GL(M, \mathbb{C})$ quotient of the holomorphic
F-term constraint, together with the King stability condition.
Therefore, we can always bring the superpotential to the form we
want, so that (\ref{eq:qs}) holds.

\subsection{Seiberg duality}

Note that given one of the above quivers, all others can be
constructed from it simply by repeated application of the rules
developed by Berenstein and Douglas \cite{BD} for generalized
Seiberg dualities. Suppose we begin with quiver $Q_n$. Then
dualizing the ${\cal O}_C(-n-1)[1]$ node reverses the directions
of all arrows through that node. In addition, between the other
two nodes there will be $2n$ new mesonic fields $M_{a i} = A_a
q_i$ for $a=1,2$, $i=1, ..., n$.

The superpotential calculated above implies that there is a mass
term $M_{a i} p_j$, where the indices are contracted as described
above to be consistent with the $SU(2)$ flavor symmetry. This
lifts all $p_j$ and a corresponding $n-1$ of the mesons from the
massless spectrum. Therefore, we obtain exactly the field content
of the quiver $Q_{n+1}$! This result had to be true, given the
previous calculation of the quiver directly from the new basis of
objects in the derived category.

\section{$\theta$ Stability and Cyclicity}
The moduli space of supersymmetric Higgs branch vacua of the
quiver quantum mechanics\footnote{Note that our quiver must be
understood as a quantum mechanics, describing the BPS
configurations of a point-like object in the $R^{3,1}$, rather than
a $3+1$ field theory, as it would then be anomalous. This is obvious
from the presence of a Calabi-Yau wrapping brane.} describing the
D6/D2/D0 bound states is given by the $U(N) \times U(M) \times
U(1)$ K\"ahler quotient of the solution of the F-flatness
conditions. The background values of the K\"ahler moduli are encoded
in the values of the FI parameters in the K\"ahler quotient.

In general, it is a very difficult problem to determine the Euler
character of the resulting moduli space, even by using the
toric action to reduce to fixed points. We will find that for a
particular choice of FI parameters the situation is dramatically
simpler. This motivates us to chooce a convenient basis of branes
(that is, a particular mutation, $Q_n$, of the quiver) for which
the FI parameters are of this simple type in a given chamber in
the background K\"ahler moduli space.

It was shown by King \cite{King} that is it possible to replace
the D-term equation appearing in the K\"ahler quotient by a purely
holomorphic algebraic condition, called $\theta$-stability. Let
$(\theta_v)_{v \in V}$ be the FI parameters, a set of real numbers
assigned to the nodes of the quiver, such that $\theta({\bf
N})=\sum N_v \theta_v =0$ for a given dimension vector ${\bf N}$.
Then a representation $R$ is called $\theta$-stable if for every
\textit{proper} subrepresentation $\tilde{R}$ with dimension
vector ${\tilde{\bf N}}$, $\theta({\tilde{\bf N}})$ is smaller
than $\theta({\bf N})$.

Consider the chamber in the space of FI parameters given by
%
%
\begin{eqnarray}
\label{eq:theta}
\theta_1 > 0, \  \theta_2<0, \ \theta_3<0, \ \ \  \text{for $Q_n$}; \nonumber \\
\theta_1 > 0, \  \theta_2<0, \ \theta_3<0, \ \ \  \text{for $\tilde{Q}_n$}.
\end{eqnarray}

\medskip

Our interest is in bound states with one unit of D6 charge, thus
we have that $N_1 = 1$. Then King stability is equivalent to
cyclicity, in the sense that the entire representation is
generated by a vector in $\cc$, the node associated to ${\cal
O}_X[1]$. Firstly, any such representation is King stable for this
choice of FI parameters, since any subrepresentation that includes
this node must be the entire representation, and thus the proper
subrepresentations all have $\theta({\tilde{\bf N}}) = \tilde N_2
\theta_2 + \tilde N_3 \theta_3 <0$.

Moreover, suppose that ${\bf R}$ is a King stable representation
with $N_1 = 1$. Then consider the subrepresentation, ${\tilde{\bf
R}}$, generated by the vector space $\cc$ of the D6 node. If it is
not all of ${\bf R}$, then $\tilde N_2 < N_2$ or $\tilde N_3 <
N_3$, and one has that $\theta({\tilde {\bf R}}) = \theta_1 +
\tilde N_2 \theta_2 + \tilde N_3 \theta_3 > 0$, and the
representation ${\bf R}$ must be unstable.

\section{The big picture: connecting the dots}

We would like to put together every piece of the story in this
section. First of all, we observe that the D6/D2/D0 BPS partition
function at a certain chamber, $\mathbb{C}_n$ in the K\"ahler
moduli space is the same as the pyramid partition generating
function for length $n$ ERC, after the parameter identification
(\ref{eq:identify}), \be Z_{D6/D2/D0}(u,v, C_n) = Z_{pyramid}(n;
q_0, q_1). \ee

By empirically checking the finite type pyramid partition, we also
conjecture that the BPS states partition function is identical to
the finite type pyramid partition generating function:

\be Z_{D6/D2/D0}(u,v, \tilde{C}_n) = Z_{finite}(n; q_0, q_1). \ee

Physically, given a set of brane charges, we should be able to use
the quiver theory to compute the Euler character of the moduli
space. In order to do that, we need to know how to translate the
data of chamber $C_n$ or $\tilde{C}_n$ into the FI parameters of
the corresponding quiver. This, in general, is a very difficult
task.

In the conifold case, we are in luck because we have the answer
from pyramid partition. We showed that pyramid partitions with
length $n$ ERC, as well as those of finite type, are torus fixed
points in the moduli space of vacua of a certain quiver. Using
this answer, we noticed that this quiver is Seiberg dual to the
quiver with basis $\{ \mathcal{O}_X[1], \mathcal{O}_{C},
\mathcal{O}_C(-1)[1] \}$. And in the Seiberg dual quiver, $Q_n$,
we can determine the FI parameters to reproduce the cyclicity
property. So we should keep in mind the following relation:

\begin{eqnarray}
Z_{D6/D2/D0}(u,v, C_n) &=& Z_{quiver}^{Q_0}(u,v,\theta_i^{Q_0})
\nonumber \\ &=& Z_{quiver}^{Q_n}(u,v,\theta_1^{Q_n}>0,
\theta_2^{Q_n}<0,\theta_3^{Q_n}<0),
\end{eqnarray}

\begin{eqnarray}
Z_{D6/D2/D0}(u,v, \tilde{C}_n) &=&
Z_{quiver}^{Q_0}(u,v,\tilde{\theta}_i^{Q_0}) \nonumber \\ &=&
Z_{quiver}^{\tilde{Q}_n}(u,v,\theta_1^{\tilde{Q}_n}>0,
\theta_2^{\tilde{Q}_n}<0,\theta_3^{\tilde{Q}_n}<0).
\end{eqnarray}

The quivers $Q_n$ and $\tilde{Q}_n$ are Seiberg dual to the quiver
$Q_0$, which has basis
$$\{ \mathcal{O}_X[1],\ \mathcal{O}_{C},\ \mathcal{O}_C(-1)[1] \}.$$


Presumably, we should be able to find the mapping: \be
\label{eq:mapping} C_n \leftrightarrow \theta_i^{Q_0}
\leftrightarrow \theta_i^{Q_n}, \ \ \tilde{C}_n \leftrightarrow
\tilde{\theta}_i^{Q_0} \leftrightarrow \theta_i^{\tilde{Q}_n}. \ee

For $Q_n$, the $\theta$ stability condition for
$\{\theta_1^{Q_n}>0, \theta_2^{Q_n}<0,\theta_3^{Q_n}<0\}$ gives
exactly the rules for constructing the pyramid partition in length
n ERC. On the other hand, the $\theta$ stability of the quiver
$\tilde{Q}_n$ with $\{\theta_1^{\tilde{Q}_n}>0,
\theta_2^{\tilde{Q}_n}<0,\theta_3^{\tilde{Q}_n}<0\}$ gives the
rules for constructing the finite type pyramid partition.

It is also possible to obtain the mapping (\ref{eq:mapping})
between the chambers in the space of K\"ahler moduli and the FI
parameters before matching the answers.
Consider the chamber $[{\cal W}^1_n {\cal W}^1_{n+1}]$, which we
checked corresponds to the simple choice of FI parameters for the
quiver $\tilde{Q}_n$ . This contains the locus $Im(z) = 0$, $-n-1<
Re(z) < -n$ along the boundary of the K\"ahler cone for $\pi/3 <
\vp < 2\pi/3$. The D2/D0 branes associated to the sheaves
$\oo_C(-n-1)[1]$ and $\oo_C(-n-2)$ have charges $-\beta + n dV$
and $+\beta - (n+1) dV$.

\begin{figure}[ht]
\centering 
\epsfig{file=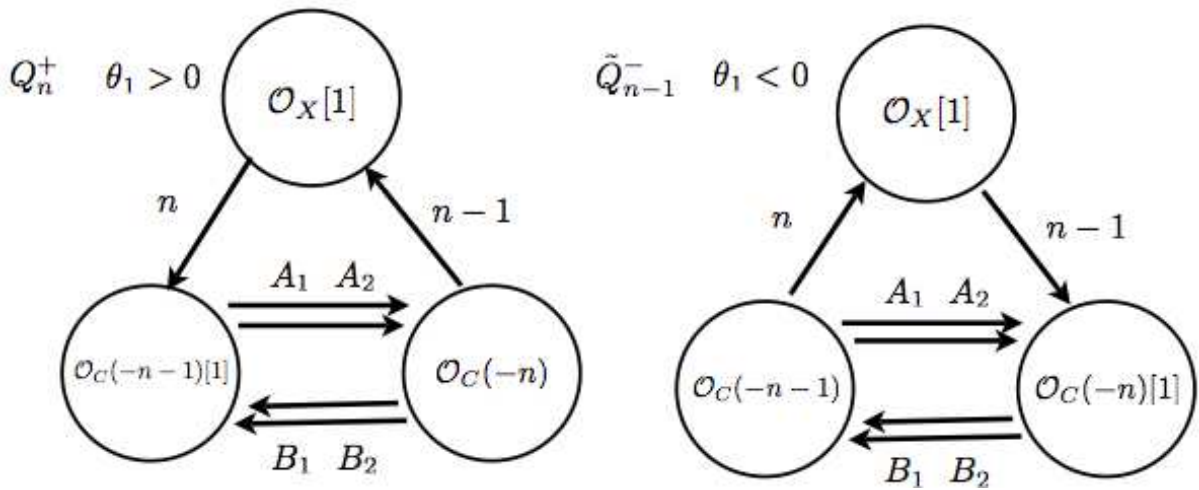, width=16cm}
\caption{The quiver $Q_n^{+}$ and $\tilde{Q}_n^{-}$.}
\label{fig:qnplus}
\end{figure}

\begin{figure}[ht]
\centering 
\epsfig{file=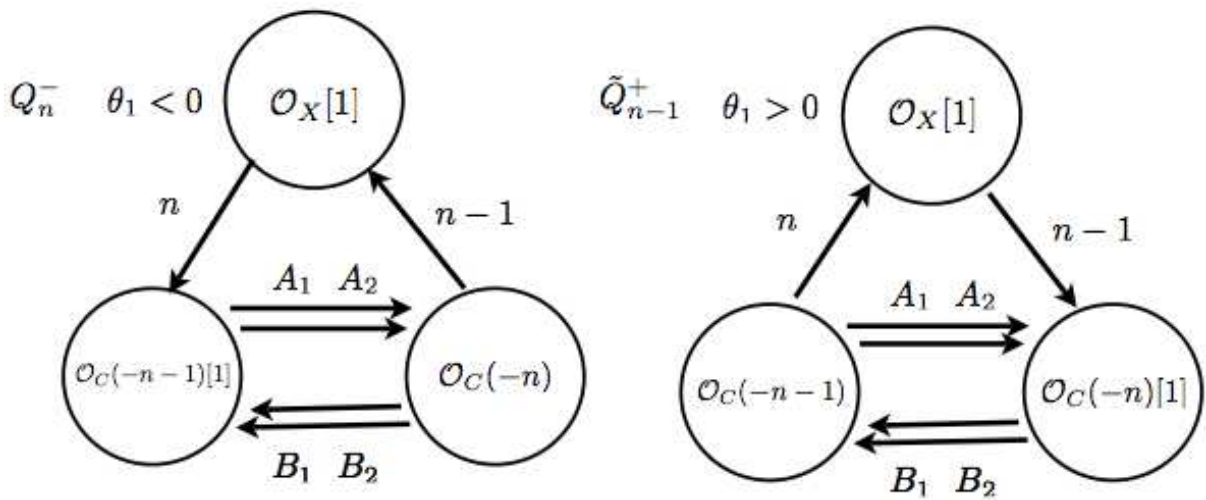, width=16cm}
\caption{The quiver $Q_n^{-}$ and $\tilde{Q}_n^{+}$.}
\label{fig:qnminus}
\end{figure}


When their central charges $Z(-\beta + n dV; t) = -z-n$ and
$Z(+\beta - (n+1) dV; t) = z + n+1$ are aligned, then the
bifundamental strings stretched between these two branes must
become massless (at tree level) in the quiver quantum mechanics.
Referring to the form of the bosonic potential, we see this occurs
precisely when the FI parameters for those nodes are equal.
Therefore we are in the chamber expected. This provides an {\it a
priori} derivation of the partition function of D6/D2/D0 bound
states in each chamber.

One last thing to notice is that we can flip the signs of the
$\theta$s and the directions of the arrows of the quiver at the
same time, without causing any change to the partition function of
the quiver theory. The reason is that in this way we do not change
the D-term conditions at all. Therefore, we have:

\be Z_{Q_n^+}(u,v)=Z_{\tilde{Q}_{n-1}^-}(u^{-1},v), \ \ \
Z_{Q_n^-}(u,v)=Z_{\tilde{Q}_{n-1}^+}(u^{-1},v) \ee where we
simplify our notation by specifying the signs of $\theta_1$ in the
quiver by putting a superscript on the $Q_n$ or $\tilde{Q}_n$.

\section{Conclusion}

In this paper we have studied the relation between the pyramid
partition generating function and the D6/D2/D0 BPS state partition
function on the resolved conifold. We found that the generating
function of the pyramid partition with length $n$ ERC is
equivalent to D6/D2/D0 BPS state partition functions in certain
chambers in the K\"ahler moduli space. More precisely, we have the
following relation:

\be Z_{D6/D2/D0}(u,v, C_n) = Z_{pyramid}(n; q_0, q_1), \ \ C_n =
[{\cal W}^{-1}_n {\cal W}^{-1}_{n-1}], \ee \be Z_{D6/D2/D0}(u,v,
\tilde{C}_n) = Z_{finite}(n; q_0, q_1), \ \ \tilde{C}_n = [{\cal
W}^{1}_{n-1} {\cal W}^{1}_{n}], \ee where the chambers $[{\cal
W}^{-1}_n {\cal W}^{-1}_{n-1}]$ and $[{\cal W}^{1}_{n-1} {\cal
W}^{1}_{n}]$ are defined in \cite{JM}.

From the rules specifying pyramid partitions (of both infinite and
finite type), we constructed the corresponding quivers, the
$\theta$s parameters, and the superpotentials. We gave the
underlying basis of sheaves and verified that they are Seiberg
dual to the original D6/D2/D0 systems. The arrow structures of the
quivers are also verified by computing $Ext$ groups. The $\theta$
parameters in these particular basis are simple and the
superpotentials are quartic, so that the rules of pyramid
partition emerge. We also noted that the cyclicity condition on
quiver representations is the same as the King stability condition
in the region of (\ref{eq:theta}).

It would also be interesting to see if there is a similar story
for the BPS partition function in the chambers other than $[{\cal
W}^{-1}_n {\cal W}^{-1}_{n-1}]$ and $[{\cal W}^{1}_{n-1} {\cal
W}^{1}_{n}]$. We also note that pyramid partitions with more
colors have been developed in \cite{BenYoung2}, and we suspect a
similar story will emerge in the case of orbifold Donaldson-Thomas
partition function.

\medskip

{\bf Acknowledgments:} We would like to thank Emanuel Diaconescu,
Greg Moore, Bal\'{a}zs Szendr\H{o}i and Alessandro Tomasiello for
useful conversations. We are also grateful to the Simons Workshop
in Mathematics and Physics 2008 for providing a stimulating
atmosphere during the final stage of this project. WYC and DJ are
supported by DOE grant DE-FG02-96ER40959.



\begin{thebibliography}{99}

\bibitem{DM}
F.~Denef and G.~W.~Moore,
  ``Split states, entropy enigmas, holes and halos,''
  arXiv:hep-th/0702146.

\bibitem{KS}
M. Kontsevich and Y. Soibelman, ``Stability structures, motivic Donaldson-Thomas invariants and cluster transformations,'' to appear.

\bibitem{qfoam}
  A.~Iqbal, N.~Nekrasov, A.~Okounkov and C.~Vafa,
  ``Quantum foam and topological strings,''
  arXiv:hep-th/0312022.

\bibitem{MNOP}
  D.~Maulik, N.~Nekrasov, A.~Okounkov, and R.~Pandharipande,
  ``Gromov-Witten theory and Donaldson-Thomas theory, I,''
  arXiv:math/0312059 ; \\
  ``Gromov-Witten theory and Donaldson-Thomas theory, II,''
  arXiv:math/0406092

\bibitem{Szendroi}
  B.~Szendr\H{o}i,
  ``Non-commutative Donaldson-Thomas theory and the conifold,''
  arXiv:0705.3419 [math.AG].

\bibitem{BenYoung}
  B.~Young, ``Computing a pyramid partition generating function with dimer shuffling,''
  arXiv:0709.3079

\bibitem{JM}
D.~L.~Jafferis and G.~W.~Moore,
  ``Wall crossing in local Calabi Yau manifolds,''
  arXiv:0810.4909 [hep-th].
  
\bibitem{KW}
  I.~R.~Klebanov and E.~Witten,
  ``Superconformal field theory on threebranes at a Calabi-Yau  singularity,''
  Nucl.\ Phys.\  B {\bf 536}, 199 (1998)
  [arXiv:hep-th/9807080].

\bibitem{Jafferis:2007sg}
  D.~L.~Jafferis,
  ``Topological Quiver Matrix Models and Quantum Foam,''
  arXiv:0705.2250 [hep-th].


\bibitem{King}
A. King, ``Moduli of representations of finite-dimensional algebras,'' Quart. J. Math. Oxford Ser. {\bf 2} 45 (1994), no. 180, 515–530.

\bibitem{Denef1}
  F.~Denef,
  ``Supergravity flows and D-brane stability,''
  JHEP {\bf 0008}, 050 (2000)
  [arXiv:hep-th/0005049].

\bibitem{Ginzburg}
 V.~Ginzburg,
  ``Calabi-Yau algebras,''
  arXiv:math/0612139.

\bibitem{AK}
  P.~S.~Aspinwall and S.~H.~Katz,
  ``Computation of superpotentials for D-Branes,''
  Commun.\ Math.\ Phys.\  {\bf 264}, 227 (2006)
  [arXiv:hep-th/0412209].

\bibitem{GH}
P. Griffiths and J. Harris, ``Principles of Algebraic Geometry,'' Wiley-Interscience, 1978.

\bibitem{Hart}
R. Hartshorne.``Algebraic Geometry,'' Springer.

\bibitem{BD}
  D.~Berenstein and M.~R.~Douglas,
  ``Seiberg duality for quiver gauge theories,''
  arXiv:hep-th/0207027.

\bibitem{BenYoung2}
 B.~Young, with an appendix by J.~Bryan, ``Generating functions for colored 3D Young diagrams and the Donaldson-Thomas invariants of orbifolds,'' arXiv:0802.3948.

\bibitem{Naka}
K.~Nagao and H,~Nakajima, ``Counting invariant of perverse coherent sheaves and its wall-corssing,''
arXiv:0809.2992.

\bibitem{Nagao}
  K.~Nagao,
  ``Derived categories of small toric Calabi-Yau 3-folds and counting invariants,''
  arXiv:0809.2994.

\end{thebibliography}
\end{document}